\documentclass[prl,twocolumn,superscriptaddress,showpacs]{revtex4}
\usepackage{graphicx,dcolumn,bm,amsmath,amssymb,color}

\begin{document}

\title{First direct mass measurement of the 
two-neutron halo nucleus $^{6}$He\\ and improved mass for the
four-neutron halo $^{8}$He}

\author{M.~Brodeur}
\email[Corresponding author:~]{brodeur@nscl.msu.edu}
\affiliation{TRIUMF, 4004 Wesbrook Mall, Vancouver BC, Canada V6T 2A3}
\affiliation{Department of Physics and Astronomy,
University of British Columbia, Vancouver, BC, Canada V6T 1Z1}
\author{T.~Brunner}
\affiliation{TRIUMF, 4004 Wesbrook Mall, Vancouver BC, Canada V6T 2A3}
\affiliation{Physik Department E12, Technische Universit\"{a}t
M\"{u}nchen, James Franck Str., Garching, Germany}
\author{C.~Champagne}
\affiliation{TRIUMF, 4004 Wesbrook Mall, Vancouver BC, Canada V6T 2A3}
\affiliation{Department of Physics, McGill University,
Montr\'{e}al, Qu\'{e}bec, Canada H3A 2T8}
\author{S.~Ettenauer}
\affiliation{TRIUMF, 4004 Wesbrook Mall, Vancouver BC, Canada V6T 2A3}
\affiliation{Department of Physics and Astronomy,
University of British Columbia, Vancouver, BC, Canada V6T 1Z1}
\author{M.J.~Smith}
\affiliation{TRIUMF, 4004 Wesbrook Mall, Vancouver BC, Canada V6T 2A3}
\affiliation{Department of Physics and Astronomy,
University of British Columbia, Vancouver, BC, Canada V6T 1Z1}
\author{A.~Lapierre}
\affiliation{TRIUMF, 4004 Wesbrook Mall, Vancouver BC, Canada V6T 2A3}
\author{R.~Ringle}
\affiliation{TRIUMF, 4004 Wesbrook Mall, Vancouver BC, Canada V6T 2A3}
\author{V.L.~Ryjkov}
\affiliation{TRIUMF, 4004 Wesbrook Mall, Vancouver BC, Canada V6T 2A3}
\author{S.~Bacca}
\affiliation{TRIUMF, 4004 Wesbrook Mall, Vancouver BC, Canada V6T 2A3}
\author{P.~Delheij}
\affiliation{TRIUMF, 4004 Wesbrook Mall, Vancouver BC, Canada V6T 2A3}
\author{G.W.F.~Drake}
\affiliation{University of Windsor, Windsor ON, Canada}
\author{D.~Lunney}
\affiliation{CSNSM-IN2P3-CNRS, Universit\'{e} Paris 11, 91405
Orsay, France}
\author{A.~Schwenk}
\affiliation{ExtreMe Matter Institute EMMI, GSI Helmholtzzentrum f\"ur
Schwerionenforschung GmbH, 64291 Darmstadt, Germany}
\affiliation{Institut f\"ur Kernphysik, Technische Universit\"at
Darmstadt, 64289 Darmstadt, Germany}
\author{J.~Dilling}
\affiliation{TRIUMF, 4004 Wesbrook Mall, Vancouver BC, Canada V6T 2A3}
\affiliation{Department of Physics and Astronomy,
University of British Columbia, Vancouver, BC, Canada V6T 1Z1}

\begin{abstract}
The first direct mass-measurement of $^{6}$He has been performed 
with the TITAN Penning trap mass spectrometer at the ISAC facility. 
In addition, the mass of $^{8}$He was determined with improved
precision over our previous measurement. The obtained masses are $m$($^{6}$He) = 6.018
885 883(57)~u and $m$($^{8}$He) = 8.033 934 44(11)~u. The $^{6}$He
value shows a deviation from the literature of 4$\sigma$. With these new mass
values and the previously measured atomic isotope shifts we obtain charge radii of
2.060(8) fm and 1.959(16) fm for $^{6}$He and $^{8}$He respectively. We present 
a detailed comparison to nuclear theory for $^6$He,
including new hyperspherical harmonics results. A correlation plot
of the point-proton radius with the two-neutron separation energy
demonstrates clearly the importance of three-nucleon forces.
\end{abstract}

\pacs{
 21.10.Dr 
 27.20.+n 
 21.45.-v 
}

\maketitle

Nuclei with exceptionally weak binding lie at the limits of stability
and exhibit fascinating phenomena. One of them is the formation of a
halo structure of one or more loosely-bound nucleons surrounding a
tightly bound core, similar to electrons in atoms. The experimentally
best studied cases are the two-neutron halo nuclei $^{6}$He and
$^{11}$Li~\cite{Tan96}. These nuclei are of Borromean nature, where
all two-body (two-neutron and neutron-core) subsystems are unbound,
but the three-body system is loosely bound~\cite{Jon04}.  
Due to a lack of pairing correlations, the neighbouring isotopes of
$^6$He ($^5$He and $^7$He) are unbound, while $^{8}$He is again bound
with a four-neutron halo. This heaviest helium isotope also marks the
nucleus with the most extreme neutron-to-proton ratio ($N/Z$ = 3).
Neutron halo nuclei are distinguished by their extended matter radius
and a small neutron separation energy compared to other nuclei. The
size of their core can be associated with the (root-mean-square)
charge radius (its deviation from the halo-free core results from
polarization effects due to strong interactions), while the halo
extension depends exponentially on the separation energy~\cite{Han87}.

To date, charge radii of halo nuclei can be determined only
from the measurement of the change in energy of an atomic transition
between isotopes $A$ and $A'$. This so-called isotopic shift
$\delta\nu^{A,A'}$ is linked to the mean-square charge radius
difference \((r_c^2)^{A} - (r_c^2)^{A'}\) by:
\begin{equation}
\delta\nu^{A,A'} = \delta\nu_{MS}^{A,A'} 
+ K_{FS} \cdot ((r_c^2)^{A} - (r_c^2)^{A'}) \,,
\label{eq:rc}
\end{equation}
where the mass shift $\delta\nu_{MS}^{A,A'}$ and the field shift constant
 $K_{FS}$ are obtained using atomic structure
calculations~\cite{Dra04}. Both terms need to be known with the same
absolute precision.  Because of their larger fractional change in mass
and their smaller volume, light nuclei have a mass shift term
typically $> 10^{4}$ times larger than the field shift
\(\delta\nu_{FS}^{A,A'} = K_{FS} \cdot ((r_c^2)^{A} - (r_c^2)^{A'})\).
Also, the mass shift depends sensitively on the nuclear mass and
in order for the mass uncertainty to make a negligible contribution to
the charge radius determination of halo nuclei, reliable atomic masses
with relative uncertainty on the order of $10^{-7}$ are needed, as for example achieved in \cite{Neo11}.

The nuclear charge radii of $^{6,8}$He have been measured by laser
spectroscopy~\cite{Wan04,Mue07}. However, to date, the mass of
$^{6}$He~\cite{Aud03} is determined only from the $Q$-value comparison
of two nuclear reactions~\cite{Rob78} and has never been measured
directly.  Over the past years, direct Penning-trap mass measurements
have uncovered large deviations with indirectly measured masses, while
yielding consistent results with other direct mass measurement methods (e.g.,
the 5$\sigma$ deviation of the $^{11}$Li mass~\cite{Smi08}).
Hence, a precise and accurate mass measurement of $^{6}$He 
is highly desirable to update the charge radius analysis.

Understanding and predicting the properties of halo nuclei also
presents a theoretical challenge. $^{6}$He and $^{8}$He are the
lightest known halo nuclei and, due to their few-nucleon ($A \lesssim
10$) structure, are amenable to different ab-initio calculations based
on microscopic nuclear forces \cite{GFMC,NCSM,FMD,MCM,EIHH}. Therefore, they represent an ideal
testing ground for nuclear structure theory, leading to a deeper
understanding of the strong interaction in neutron-rich systems.

In this Letter, we present the first direct mass measurement of
$^{6}$He, together with a more precise value for $^{8}$He, using the
TRIUMF Ion Trap for Atomic and Nuclear science (TITAN)~\cite{Dil06}
Penning trap mass spectrometer. The TITAN facility is situated in the
low-energy section of the TRIUMF's Isotope Separator and ACcelerator
(ISAC) experimental hall~\cite{Dom00}. The $^{8}$He mass was first
directly measured in an earlier TITAN experiment~\cite{Ryj08}. Based
on the new masses presented here, we determine reliable binding
energies, and the resulting values for the charge radii $r_{\rm c}$ of
$^{6}$He and $^{8}$He. These observables provide key tests for nuclear
theory. We make a detailed comparison to theory for $^6$He, where
ab-initio calculations based on different nucleon-nucleon (NN) and
three-nucleon (3N) forces are available.
To date, no calculation exists based on chiral effective field theory
interactions. This approach has the advantage that the corresponding
3N and 4N forces are largely predicted.
As a first step towards this goal, we present new
ab-initio hyperspherical harmonics results based on chiral
low-momentum interactions. A natural correlation between separation energy and
radii is found when only NN interactions are included. The results and
the precise experimental data clearly illustrate the importance of
including 3N forces.

Both radioactive helium isotopes were produced via spallation reaction
using 500 MeV protons from the TRIUMF cyclotron at a current of 80
$\mu$A impinging a high power silicon-carbide target. The beam was
ionized using the Forced Electron Beam Ion Arc Discharge (FEBIAD)
source~\cite{Bri08} and transported at an energy of 20 keV to the
TITAN facility. Contamination in both beams was removed using a
two-stage high resolving power dipole-magnet mass separator. Upon
reaching the TITAN facility, the purified continuous ion beam was
thermalized, accumulated and bunched using a hydrogen-filled Radio
Frequency Quadrupolar (RFQ) ion trap~\cite{Smi06}. After their
extraction from the RFQ, the ions were transported at an energy of
approximately one keV to the Penning trap, where the mass measurement
was performed.

The basic principle behind Penning trap mass spectrometry consists of
measuring the cyclotron frequency \(\nu_{c} = qB/(2 \pi M)\) of an ion
of mass $M$ and charge $q$ in a magnetic field $B$. TITAN, like most
on-line Penning trap mass spectrometers, uses the Time Of Flight
Ion-Cyclotron Resonance (TOF-ICR) technique~\cite{Gra80, Kon95} to
determine the ion's cyclotron frequency (we refer the reader
to~\cite{Bro09} for more details about mass measurements using the
TOF-ICR technique at TITAN).

\begin{figure}[t]
\begin{center}
\includegraphics[width=0.49\textwidth,clip=]{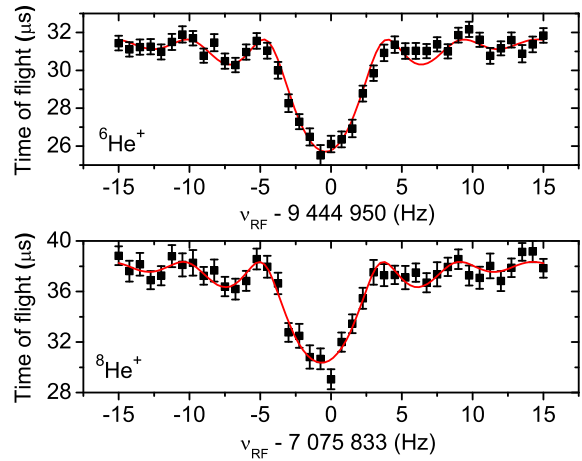}
\caption{(Color online) Time-of-flight resonance spectra of 
$^{6}$He$^{+}$ and $^{8}$He$^{+}$. The solid line (red) is a fit
of the theoretical line shape~\cite{Kon95}. The shortest 
time-of-flight is achieved when the ions are excited at the 
cyclotron frequency, i.e., when \(\nu_{RF} = \nu_{c}\).
\label{fig:He6He8tspec}}
\end{center}
\end{figure}

Typical $^{6}$He$^{+}$ and $^{8}$He$^{+}$ time-of-flight ion-cyclotron
resonances are shown in Fig.~\ref{fig:He6He8tspec}. These measurements
took 27 minutes each and comprised 1656 and 1171 detected ions
yielding statistical relative uncertainties on the cyclotron
frequencies of 9 and 14 parts per billion (ppb), respectively. For
both isotopes, the magnetic field was calibrated by measuring the
cyclotron frequency of stable $^{7}$Li$^{+}$ produced by the TITAN
off-line ion source between the $^{6}$He$^{+}$ (or $^{8}$He$^{+}$)
cyclotron frequency measurements. From these measurements, one
calculates the frequency ratio \(R =
\nu_{c}(^{7}\mbox{Li}^{+})/\nu_{c}(^{6,8}\mbox{He}^{+})\), which
yields the ratio of the masses of the two ions.




A total of 12 $^{6}$He$^{+}$ and 17 $^{8}$He$^{+}$ frequency ratios
where measured and for these measurements, the different sources of
systematic errors such as magnetic field inhomogeneities, misalignment
with the magnetic field, harmonic distortion of the trap potential,
non-harmonic terms in the trapping potential, interaction of multiple
ions in the trap, magnetic field time-fluctuations and error due to
relativistic effects were considered (see~\cite{Bro10} for a detailed
analysis and treatment of these effects for the $^{6,8}$He
measurements). The main systematic errors on the $^{6}$He
and $^{8}$He cyclotron frequency ratios arise from the interaction of
multiple ions in the trap and are found to be 8 and 13 ppb for
$^{6}$He$^{+}$ and $^{8}$He$^{+}$, respectively. The contributions from
the other effects are all below the ppb level and therefore have a
negligible contribution to the final uncertainty. The weighted
averages of the cyclotron frequency ratios $\overline{R}$ are 0.857
868 442 9(42)$\{$82$\}$ and 1.145 098 361(7)$\{$16$\}$ for $^{6}$He
and $^{8}$He, respectively (where the statistical uncertainty is given
in parenthesis and the total uncertainty in curly brackets).

In mass spectrometry, the quantity of interest is the atomic mass,
which is given by \(m = \overline{R} \cdot (m_{cal} - m_{e} +
B_{e,cal}) + m_{e} - B_{e}\), where $B_{e,cal}$ and $B_{e}$ are the
last electron binding energies of the calibrant ion and of the ion of
interest, $m_{e}$ is the electron mass, and $m_{cal}$ is the calibrant
atomic mass.

Using the more precise mass measurement of the calibrant $^{6}$Li
from~\cite{Mou10}, the $^{8}$He mass reported in~\cite{Ryj08} becomes
8.033 935 67(72) u. The $^{8}$He measurement presented here yields a
mass of 8.033 934 40(12) u, which agrees with the previous result
within 1.7$\sigma$, with a factor of 12 improvement in precision. Combining these
two results, the mass and mass excess of $^{8}$He become 8.033 934
44(11) u and 31 609.72(11) keV. This is within 1.7$\sigma$ of the
atomic mass evaluation (AME03) value~\cite{Aud03}. On the other hand,
for the $^{6}$He mass and mass excess we obtain 6.018 885 883(57) u
and 17 592.087(54) keV, which deviate from the AME03 by 4$\sigma$
while improving the precision by a factor of 14.

\begin{table}[t]
\begin{center}
\begin{tabular}{|c|c|c|c|}
\hline
Transition & $\delta\nu^{A,4}$ & $\delta\nu_{MS}^{A,4}$ & $\delta\nu_{FS}^{A,4}$ \\
\hline \hline
$^{8}$He 2$^{3}$S$_{1}$ $\rightarrow$ 3$^{3}$P$_{1}$ & 64701.129(73) & 64702.0982 & -0.969(73) \\
\hline
$^{8}$He 2$^{3}$S$_{1}$ $\rightarrow$ 3$^{3}$P$_{2}$ & 64701.466(52) & 64702.5086 & -1.043(52) \\
\hline
mean + nucl. pol. & & & -1.020(42)\{64\} \\
\hline \hline
$^{6}$He 2$^{3}$S$_{1}$ $\rightarrow$ 3$^{3}$P$_{0}$ & 43194.740(37) & 43196.1573 & -1.417(37) \\
\hline
$^{6}$He 2$^{3}$S$_{1}$ $\rightarrow$ 3$^{3}$P$_{1}$ & 43194.483(12) & 43195.8966 & -1.414(12) \\
\hline
$^{6}$He 2$^{3}$S$_{1}$ $\rightarrow$ 3$^{3}$P$_{2}$ & 43194.751(10) & 43196.1706 & -1.420(10) \\
\hline
mean + nucl. pol. & & & -1.431(8)\{8\} \\
\hline
$^{6}$He 2$^{3}$S$_{1}$ $\rightarrow$ 3$^{3}$P$_{2}$ & 43194.772(33) & 43196.1706 & -1.399(33)\{40\} \\
\hline 
mean & & & -1.430(8)\{31\} \\
\hline 
\end{tabular}
\caption{Isotopic shift values from~\cite{Mue07} (except the last 
transition, which is from~\cite{Wan04}), together with the
new calculated mass shifts $\delta\nu_{MS}^{A,4}$ and the new 
field shift for $^{6,8}$He using the masses measured by the TITAN 
Penning trap spectrometer. ``Mean + nucl. pol.'' gives the weighted mean of the 
transitions presented above plus the nuclear polarization correction. 
Statistical error given in ( ) and the total error in \{ \}. The error on the $^{6,8}$He 
mass shifts are 0.8 and 0.9 kHz respectively. All units are in MHz.\label{tab:NewMassShift}}
\end{center}
\end{table}

Following the TITAN measurements, the new $^{6}$He and $^{8}$He
two-neutron separation energies are 975.46(23)~keV and
2125.00(33)~keV, respectively. Using the new masses we also computed
the charge radii of $^{6,8}$He following the procedure presented
in~\cite{Mue07}. The TITAN masses enters in the mass shift
$\delta\nu_{MS}^{A,4}$ evaluation obtained from atomic structure
calculations~\cite{Dra04}. These new mass shifts, together with the
corresponding isotopic shifts $\delta\nu^{A,4}$
from~\cite{Mue07,Wan04} and updated field shifts
$\delta\nu_{FS}^{A,4}$ are presented in Table~\ref{tab:NewMassShift}.
The total field shift for $^{8}$He was taken as the weighted average
of all transitions and the various systematic uncertainties presented
in~\cite{Mue07} were added in quadrature yielding a field shift of
-1.020(64) MHz. For $^{6}$He,~\cite{Wan04} and~\cite{Mue07} were
treated as independent measurements. Consequently, we took the
weighted average of the two final field shifts, except for the
Zeeman systematic uncertainty (0.03 MHz) which was present in
both measurements. We also applied the nuclear polarization correction
(-0.014(3) MHz) to the measurement~\cite{Wan04} as done
in~\cite{Mue07}. The total field shift for $^{6}$He is then -1.430(31)
MHz. The resulting mean-square charge radii $(r_{c}^{2})^{A}$ of
$^{6,8}$He are computed using Eq.~\eqref{eq:rc}, where
$(r_{c})^{A'=4}$ = 1.681(4) fm~\cite{Sic08} is the mean-square charge
radius of $^{4}$He, and $K_{FS}$ = 1.008
MHz/fm$^{2}$~\cite{Dra04}. The updated values for the $^{6,8}$He
charge radii are 2.060(8) fm and 1.959(16) fm, respectively. 
The new mass measurements lead to a decrease in the $^{6}$He charge radius by
0.011 fm and an increase in $^{8}$He by 0.025 fm compared to the
values of~\cite{Mue07} with the $^{4}$He charge radius from~\cite{Sic08}, 
which significantly reduces the difference between the two isotopes.

In order to compare the
experimental charge radii with theory, we also calculate the
corresponding point-proton radii $r_{\rm pp}$ given by~\cite{Ong10}:
\begin{equation}
r^{2}_{\rm pp} = r^{2}_{\rm c} - R^{2}_{p} - (N/Z) \cdot R^{2}_{n} -
3/(4 M^{2}_{p}) - r^{2}_{\rm so}\,,
\label{rpp}
\end{equation}
where $R^{2}_{p}$ and $R^{2}_{n} = -0.1161(22)$ fm$^{2}$~\cite{Nak10}
are the proton and neutron mean-square charge radii, respectively,
$3/(4 M^{2}_{p}) = 0.033$ fm$^{2}$ is a first-order relativistic
(Darwin-Foldy) correction~\cite{Fri97} and $r^{2}_{\rm so}$ is a
spin-orbit nuclear charge-density correction.  The latter is estimated
to be $-0.08$ fm$^2$ and $-0.17$ fm$^2$ in the extreme case of pure
$p$-wave halo neutrons~\cite{Ong10} for $^{6}$He and $^{8}$He,
respectively (see also~\cite{Papa11} for an improved
estimate). Realistic values should be somewhere between zero and these
extremes, so we conservatively took 0.08 and 0.017 fm$^2$ as the
corresponding error.

\begin{figure}[t]
\begin{center}
\includegraphics[width=0.49\textwidth,clip=]{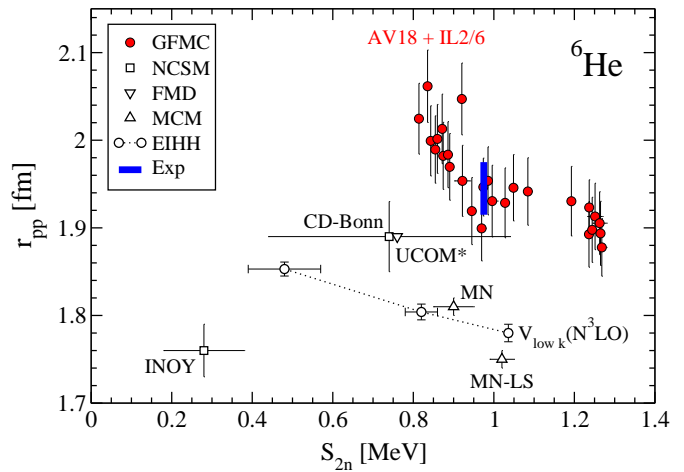}
\caption{(Color online) Correlation plot of the $^6$He point-proton
radius $r_{\rm pp}$ versus two-neutron separation energy
$S_{2n}$. The experimental range (bar) is compared to theory based
on different ab-initio methods using different NN interactions only
(open symbols) and including 3N forces fit to light nuclei (filled
symbols). The ab-initio methods are indicated in the legend and the
nuclear forces next to the symbols (for details see text).
Theoretical error bars are shown where available.\label{fig:He6_Sn_rpp}}
\end{center}
\end{figure}


For $R_{p}$ the Review of Particle Physics~\cite{Nak10} value is
$0.877(7)$ fm. Recently, $R_{p}$ has been also precisely measured from
spectroscopy of muonic hydrogen~\cite{Poh10} leading to $0.84184(67)$
fm.  Using these two values for $R_{p}$ with the above mentioned
spin-orbit corrections in Eq.~({\ref{rpp}}) we obtain $r_{\rm
pp}=1.938 \pm 0.023$ ($1.885 \pm 0.048$) and $1.953 \pm 0.022$
($1.901 \pm 0.048$) fm for $^{6}$He ($^{8}$He), respectively. The
experimental range in Fig.~\ref{fig:He6_Sn_rpp} includes both cases
within the errors shown for $^6$He.

In Fig.~\ref{fig:He6_Sn_rpp}, we compare the point-proton radius and
the two-neutron separation energy $S_{2n}$ of $^6$He to ab-initio
calculations based on different NN and 3N interactions. The Green's
Function Monte Carlo (GFMC) results~\cite{GFMC} are the only existing
converged calculations that include 3N forces, which are constrained
to reproduce the properties of light nuclei, including $^6$He and
$^8$He. The scatter in Fig.~\ref{fig:He6_Sn_rpp} gives some measure of
the numerical uncertainty in the GFMC method as well as an uncertainty
in the 3N force models used (the IL2 and IL6 three-body forces were
used with the AV18 NN potential)~\cite{GFMC}. The comparison of the
experimental range to theory clearly demonstrates the importance of
including and advancing 3N forces. The theoretical results shown in
Fig.~\ref{fig:He6_Sn_rpp} based on NN interactions only are
consistently at lower $S_{2n}$ and smaller $r_{pp}$ values. The
NN-only calculations include the Fermionic Molecular Dynamics (FMD)
results based on the UCOM NN potential and a phenomenological term (to
account for three-body physics)~\cite{FMD}, the No-Core Shell Model
(NCSM) results based on the CD Bonn and INOY NN
potentials~\cite{NCSM}, and variational Microscopic Cluster Model
(MCM) results based on the Minnesota (MN) and MN without spin-orbit
(MN-LS) NN potentials~\cite{MCM}.  Figure~\ref{fig:He6_Sn_rpp} also
shows the importance of comparing theoretical predictions to more than
one observable. To illustrate this, both NCSM (using CD Bonn) and the
GFMC results show a good agreement for the point-proton radius, while
the NCSM result has a large error for $S_{2n}$ and tends to
underpredict the two-neutron separation energy.

In addition, we present new Effective Interaction Hyperspherical
Harmonics (EIHH) results~\cite{EIHH} based on chiral low-momentum NN
interactions $V_{{\rm low}\,k}$~\cite{forces}. In the EIHH approach
the wave function falls off exponentially by construction, making it
ideally suited for the study of halo nuclei (for calculational details
see~\cite{EIHH}). The obtained energies and radii are converged within
the few-body calculational uncertainty given by the error bars.  The
three EIHH results shown in Fig.~\ref{fig:He6_Sn_rpp} are for
different NN cutoff scales $\Lambda = 1.8, 2.0$, and $2.4 \, {\rm
fm}^{-1}$. The running of observables with $\Lambda$ is due to
neglected many-body forces. The EIHH results lie on a line indicated
in Fig.~\ref{fig:He6_Sn_rpp}, leading to a decreasing $S_{2n}$ and
increasing $r_{\rm pp}$, that does not go through the experimental
range. Such a correlation is expected, because a smaller $S_{2n}$
stretches out the core~\cite{Mue07}. This correlation is also similar
to the Phillips and Tjon lines in few-body systems~\cite{few}, which
arises from strong NN interactions (large scattering
lengths). Three-body physics manifests itself as a breaking from this
line/band. The correlation is also supported by the variational MCM
results. A key future step will be to include chiral 3N forces in the
EIHH calculations.


We have presented the first direct mass-measurement of the two-neutron
halo nucleus $^{6}$He and a more precise mass value for the
four-neutron halo $^{8}$He. Both measurements where performed using
the TITAN Penning trap mass spectrometer. While the $^{8}$He mass
value is 1.7$\sigma$ within the AME03~\cite{Aud03}, the $^{6}$He mass
deviates by 4$\sigma$. The new masses lead to improved values of the
charge (and point-proton) radii and the two-neutron separation
energies, which combined provide stringent tests for three-body forces
at neutron-rich extremes. 

This work was supported by the Natural Sciences and Engineering
Research Council of Canada (NSERC) and the National Research Council
of Canada (NRC). We would like to thank the TRIUMF technical staff,
especially Melvin Good. S.E.~acknowledges support from the Vanier
CGS program, T.B.~from the Evangelisches Studienwerk e.V.~Villigst,
D.L.~from TRIUMF during his sabbatical, and A.S.~from the
Helmholtz Alliance HA216/EMMI.

\end{document}